**PtSe$_2$ grown directly on polymer foil for use as a robust piezoresistive sensor**

*Conor S. Boland\*, Cormac Ó Coileáin, Stefan Wagner, John B. McManus, Conor P. Cullen, Max C. Lemme,  Georg S. Duesberg and Niall McEvoy\**

Dr. C. S. Boland, School of Mathematical and Physical Sciences, University of Sussex, Sussex BN1 9QH, UK and School of Physics and AMBER, Trinity College Dublin, Dublin 2, Ireland
Email: bolandc1@tcd.ie
Dr. C. Ó Coileáin, J. B. McManus, C. P. Cullen, Dr. N. McEvoy, School of Chemistry and AMBER, Trinity College Dublin, Dublin 2 Ireland
Email: nmcevoy@tcd.ie
Dr. S. Wagner, Prof. M. C. Lemme, AMO GmbH, Advanced Microelectronic Center Aachen (AMICA), Otto-Blumenthal-Str. 25, 52074 Aachen, Germany
Prof. M. C. Lemme, Chair of Electronic Devices, Faculty of Electrical Engineering and Information Technology, RWTH Aachen University, Otto-Blumenthal-Str. 2, 52074 Aachen, Germany
Prof. G. S. Duesberg, Faculty of Electrical Engineering and Information Technology, EIT2 Universität der Bundeswehr München, 85577 Neubiberg, Germany

Robust strain gauges are fabricated by growing PtSe$_2$ layers directly on top of flexible polyimide foils. These PtSe$_2$ layers are grown by low-temperature, thermally-assisted conversion of predeposited Pt layers. Under applied flexure the PtSe$_2$ layers show a decrease in electrical resistance signifying a negative gauge factor. The influence of the growth temperature and film thickness on the electromechanical properties of the PtSe$_2$ layers is investigated. The best-performing strain gauges fabricated have a superior gauge factor to that of commercial metal-based strain gauges. Notably, the strain gauges offer good cyclability and are very robust, surviving repeated peel tests and immersion in water. Furthermore, preliminary results indicate that the stain gauges also show potential for high-frequency operation. This host of advantageous properties, combined with the possibility of further optimization and channel patterning, indicate that PtSe$_2$ grown directly on polyimide holds great promise for future applications.

**1. Introduction**

The emergence of 2D materials has triggered a proliferation of new research topics with much focus placed on synthesizing and modifying these materials for use in a wide array of potential applications. Many members of the transition metal dichalcogenide (TMD) family possess a bandgap and so have been touted for use in applications ranging from transistors, to photodetectors, to sensors[1]. While most studies to date have focused on group-6 TMDs, such as $MoS_2$ or $WSe_2$, recent reports have outlined the practicality of growing noble-metal, or group-10, TMDs such as $PtSe_2$. Interestingly, the band structure of $PtSe_2$ is heavily dependent on its layer thickness with bulk films having semimetallic character while mono- and few-layer films are semiconducting[2-4].

Vapour-phase growth methods, such as chemical vapour deposition (CVD) and thermally assisted conversion (TAC), offer scalable growth of TMD layers with reasonable to good crystalline quality[5-6]. While $MoS_2$ and related materials synthesized in this manner have shown impressive performance, their practical use has been somewhat hampered by the associated high growth temperature (typically > 600°C). This high temperature restricts the choice of substrate and renders the growth process incompatible with standard semiconductor processes, such as back-end-of-line (BEOL) processing. On the other hand, $PtSe_2$ can be grown at a relatively low growth temperature by a TAC process[7-8]. The material grown in this manner is nanocrystalline but, despite this, studies have indicated that it holds promise for applications in photodetectors[9-10], gas sensing[8], electronics[11-13] and electrocatalysis[14]. This relatively low growth temperature means that $PtSe_2$ can be grown directly on substrates that are inaccessible to standard vapour-phase growth processes for TMDs.

Recently, there has been an emergent need for robust, sensitive and low-power strain gauges in civil engineering that can be applied to monitor stresses on wind turbine blades[15], gear boxes[16] and bases[17] in extreme conditions. Through better understanding of the forces applied, the construction of more-efficient, green-energy generating structures is made possible. Such strain gauges are generally based on materials that exhibit piezoresistive properties. Upon the application of strain, $\varepsilon$, electrical resistance changes such that

$$\Delta R / R_0 = G\varepsilon$$

$$G = \frac{1}{\rho_0}\frac{d\rho}{d\varepsilon} + (1+2\nu)$$

Where G is the sensitivity metric known as the gauge factor measured at low strain, $\rho_0$ is the material's zero-strain resistivity, $\rho$ is the resistivity and $\nu$ is the Poisson ratio.[18]

Most commercially-available sensors are metal-based foils on a rigid polymer backing that tend to have low values of G in the range of 2-4.[18] This is due to metals undergoing small changes in resistivity with applied strain. However, for many semiconducting materials applying strain leads to changes in the band structure which can modify either carrier density or mobility, resulting in large changes in resistivity and thus high values of G. For p-type silicon, G values up to ~175 have been reported.[19] However, such materials are often brittle resulting in small sensing ranges, i.e. failure occurs at low strain.

Research has turned towards materials science, and more specifically 2D materials, in an effort to overcome these challenges. Many 2D materials display piezoresistive properties suggesting they could find use as active components in assorted electromechanical sensors such as strain gauges, as well as MEMS and NEMS devices[20-22]. $MoS_2$, a semiconducting 2D material, has been shown to have negative gauge factors ranging from -225 for bilayers to -50 for few-layer samples.[21] Most recently, pressure sensors and strain gauges based on TAC-grown $PtSe_2$, which was grown on $SiO_2$/Si and subsequently transferred onto flexible substrates, have been demonstrated[23]. The fabrication process for such devices could be greatly simplified by direct growth of $PtSe_2$ on flexible substrates. Additionally, without the need for polymer-based transfer processes, cleaner interfaces and resultant improved device performance can be anticipated. Furthermore, TAC growth is compatible with standard processing and patterning techniques and so different device geometries could be patterned directly on flexible substrates.

Herein we demonstrate the growth of PtSe$_2$ layers directly on flexible polyimide foils. These films adhere strongly to the polyimide substrate and are continuous and conductive over large areas. The flexibility of polyimide means that the effect of flexure, and other strain, on the PtSe$_2$ layers can be investigated. The PtSe$_2$ layers show a strong and stable electrical response to flexure suggesting that they could be used in future electromechanical sensors.

## 2. Results and Discussion

PtSe$_2$ thin films were grown on polyimide substrates by a TAC process as depicted in **Figure 1(a)** and described in more detail in our previous reports[7-8]. Substrates were typically ~2 cm x 4 cm in area, as shown in **Figure 1(b)**, but larger or smaller substrates could readily be made. Scanning electron microscopy (SEM) and scanning probe microscopy were used to examine the topography and uniformity of the films grown. A SEM image of a PtSe$_2$ film derived from 1 nm Pt on polyimide, converted to PtSe$_2$ at 400 $^{\circ}$C, is shown in **Figure 1(c)**. This image shows film continuity over a large area and also highlights the nanocrystalline morphology, which is typical tor TAC-grown TMD films[13]. Additional SEM images at lower magnification are shown in Figure S1(a) of the Supporting Information and indicate that the PtSe$_2$ conforms to features on the polyimide surface. Atomic force microscopy (AFM) was also used to investigate the morphology of PtSe$_2$ on polyimide, prepared under the same conditions, as shown in **Figure 1(d)**. Like the SEM image, the AFM image shows the nanocrystalline nature of the PtSe$_2$. A root-mean-square (RMS) roughness of 4.6 nm was extracted for PtSe$_2$ films grown from 1 nm Pt on polyimide, which is comparable with TAC-derived PtSe$_2$ films grown on other substrates, such as SiO$_2$/Si. Conductive-AFM (or c-AFM) imaging was also performed on the PtSe$_2$ films as shown in **Figure 1(e)**. This image was taken over a large area (30 x 30 μm) and highlights the uniformity in the conductivity of the film at this scale. The chemical composition of the films grown was probed by X-ray photoelectron spectroscopy (XPS). The Pt 4f and Se 3d spectral regions for a 1 nm Pt film converted at 400 $^{\circ}$C are shown in **Figure 1(f)**. These spectra are consistent with previous reports for PtSe$_2$ grown on more conventional Si-based substrates and indicate that stoichiometric PtSe$_2$ is formed by our process[8]. The Pt 4f PtSe$_2$ component was found to be at a binding energy of ~73 eV, a minor component likely from platinum oxide was also present at ~72 eV. For the Se 3d core level spectral region the PtSe$_2$ main component was present at ~54.5 eV. A much lower concentration component at ~55.5 eV was attributed to the presence of edge/ amorphous Se in the sample. Raman spectroscopy is usually one of the first tools used to confirm the presence of PtSe$_2$ or other TMDs[7] but in the case of direct growth on polyimide it is not possible to get a strong signal.

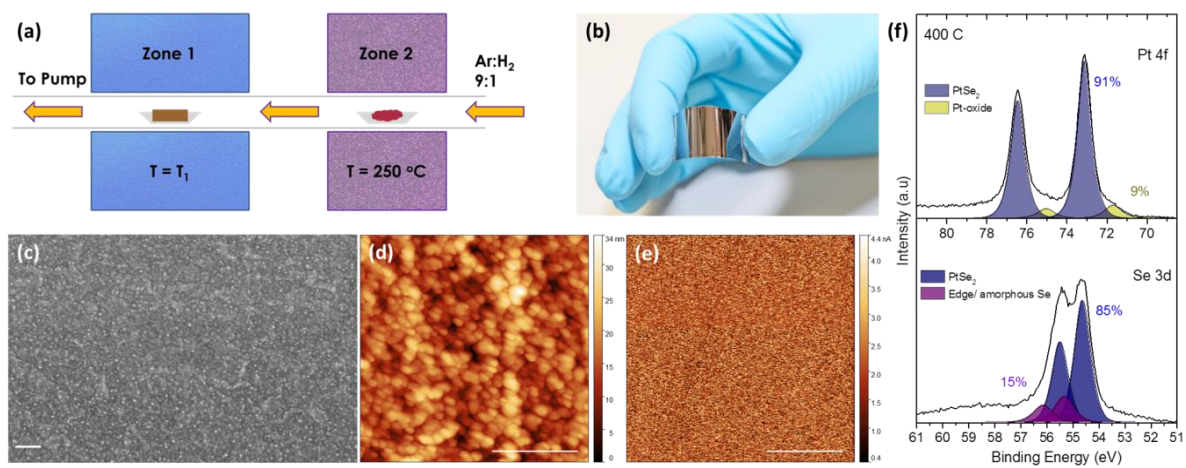

Figure 1 (a) Schematic of TAC process in two-zone furnace. Analysis of PtSe$_2$ film grown from 1 nm thick Pt on polyimide at a conversion temperature of 400 $^o$C (b) Photograph of PtSe$_2$ film grown directly on polyimide foil. (c) SEM image of PtSe$_2$ film on polyimide, scale bar is 200 nm. (d) AFM image of PtSe$_2$ film on polyimide, scale bar is 500 nm. (e) c-AFM image of PtSe$_2$ film on polyimide showing uniform conductivity over the relatively large area probed, scale bar is 10 μm. (f) Pt 4f and Se 3d core level XPS spectra of PtSe$_2$ film on polyimide.

The electromechanical properties of PtSe$_2$ on polyimide were tested by flexing the films while simultaneously measuring the films' resistance using a custom-made, three-point flexure rig, as detailed in the methods section. Under applied strain a decrease in resistance was observed, resulting in a negative gauge factor consistent with previous reports on other TMDs[21], and also with previous experimental reports on PtSe$_2$ films transferred onto a bending-beam cantilever setup[23]. In previous reports on TMDs, the negative gauge factor has been attributed to an increase in the density of states and a decrease in the size of the bandgap under applied strain[21, 23]. The effect of film thickness on the electromechanical properties was examined by converting Pt films of different starting thicknesses to $PtSe_2$ at 400 °C. Like other TMDs, the band structure and electronic properties of $PtSe_2$ depend on layer thickness; it is semiconducting in its mono- and few-layer form it is semimetallic in its bulk form[3, 7]. Plots of fractional resistance change versus applied strain, for different film thicknesses, are shown as a function of applied strain (**Figure 2(a, b)**) and flexure angle (**Figure 2(c, d)**). The labelled thicknesses on the plots refer to the starting Pt film thickness (i.e. the film thickness before TAC). It is clear from these data that the films from 1 nm starting thickness of Pt have the best properties in terms of electromechanical response and signal-to-noise ratio. This is supported by the extracted gauge factors, shown in **Figure 2(e, f)**, which are highest for the 1 nm film. The associated strain-related gauge factor is 12, which is superior to that of typical commercial metal-based strain gauges. It is likely that this performance could be improved through patterning of the channel i.e. creating a serpentine channel similar to those used in commercial strain gauges. Our previous studies indicated that conversion of 1 nm of Pt resulted in $PtSe_2$ films with a thickness of ~3.5 nm (or ~5 layers)[13]. At this layer thickness it is possible that the films are still partially semiconducting in character as there is some debate in the literature as to the exact layer thickness at which $PtSe_2$ transitions from semiconducting to semimetallic[2, 24-25]. One would perhaps expect a film of 0.5 nm Pt starting thickness to be more sensitive as the resultant $PtSe_2$ would be more semiconducting in nature[13]. However, the film uniformity and conductivity must also be considered. Films grown from 1 nm Pt were also tested in a cantilever bending-beam setup and showed comparable results to previously-investigated transferred $PtSe_2$ films as detailed in the Supporting Information, S2. It is interesting that all films display a negative gauge factor given that the thicker films are expected to be semimetallic. This is consistent with previous results on transferred $PtSe_2$ films where density functional theory calculations were used to explain the negative gauge factor in the high-strain regime[23]. Previous reports experimental reports have shown negative gauge factors for $MoS_2$[21] and positive gauge factors for graphene[26]. Whereas theoretical investigations have suggested that MoSe$_2$ and WSe$_2$ have positive gauge factors. It is clear that more detailed investigation will be required to fully understand the sign of the gauge factor for PtSe$_2$ and other 2D materials. This is particularly complicated in the case of TAC-grown PtSe$_2$ where the layer thickness has a massive impact on the electrical properties and the nanosized crystalline domains are very different to the idealized highly-crystalline scenario. Reference films of Pt only, with no selenisation treatment, show a smaller positive gauge factor when subjected to the same electromechanical tests, as detailed in the Supporting Information, Figure S3. This is the expected behavior for metallic films. Interestingly, films which are sulfurised rather than selenised also show a positive gauge factor, as shown in Figure S4 of the Supporting Information. In this case it is probable that PtS rather than PtS$_2$ is formed under the low-pressure conversion conditions used, in line with previous reports[13].

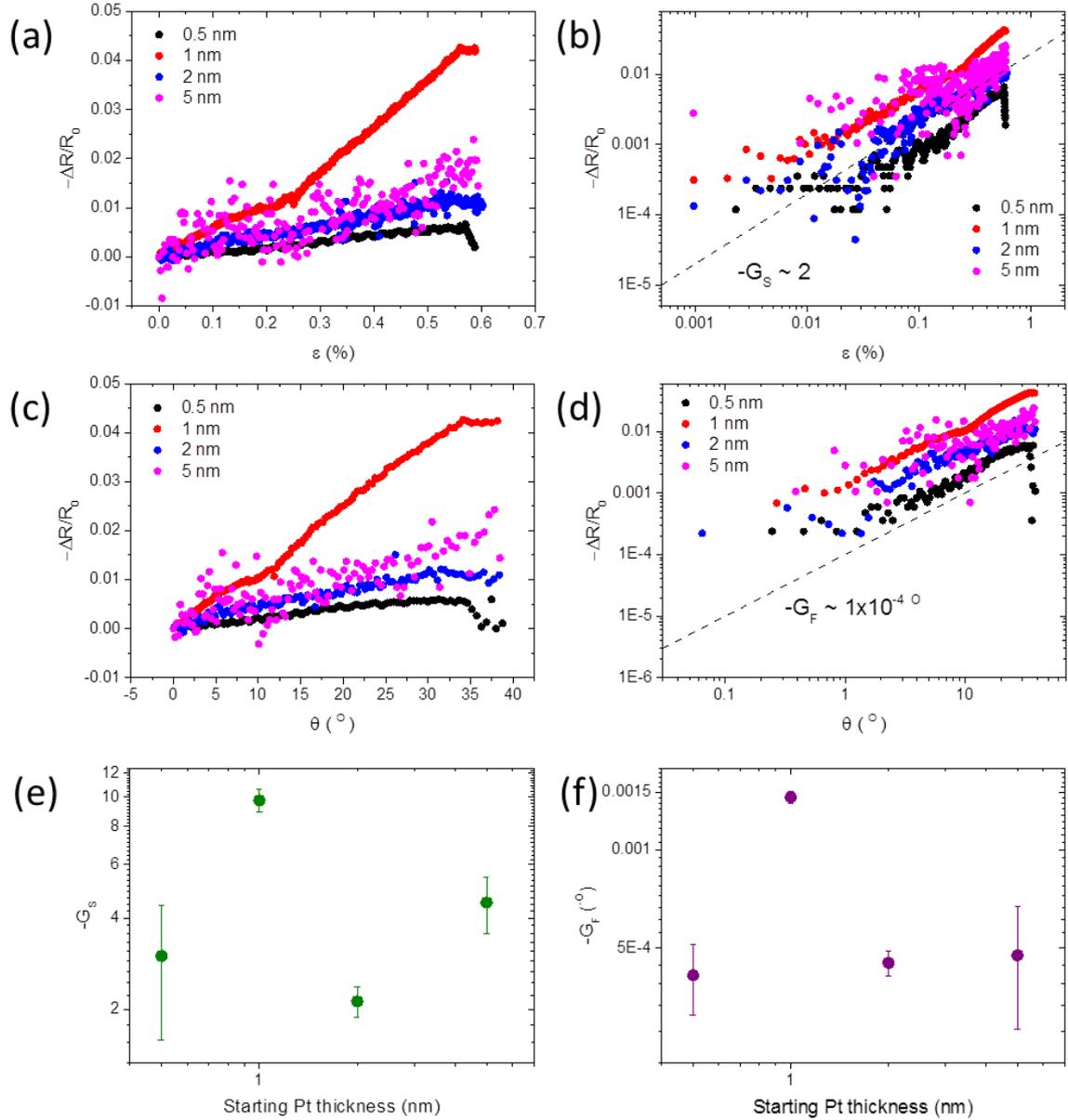

*Figure 2 Investigation of the effect of PtSe$_2$ film thickness on the electromechanical response (a, b) Fractional resistance change as a function of applied strain. (c, d) Fractional resistance change as a function of flexure angle. (e, f) Extracted gauge factors for films of different thickness.*

Our previous studies indicated that a temperature of ~400 $^{\circ}$C is required for complete conversion of Pt to PtSe$_2$. However, other reports have shown that PtSe$_2$ can be grown by epitaxial means on single-crystal Pt at temperatures as low as 270 $^{\circ}$C[3]. With this in mind, we examined the electromechanical properties of PtSe$_2$ films formed by converting 1 nm Pt at different temperatures. Shown in **Figure 3(a)** is the fractional resistance change versus strain for films converted at temperatures in the range 250-450 °C. It is worth noting that the polyimide substrate would not survive temperatures higher than this and, furthermore, PtSe$_2$ would not form. **Figure 3(a)** indicates that the strongest electromechanical response is obtained from films with a conversion temperature of 400 °C. These films have the lowest resistance and highest (negative) gauge factor as shown in **Figure 3(b)** and **3(c)**, respectively. This suggests that complete conversion of Pt to PtSe$_2$ is required for a strong electromechanical response – although it is also possible that the crystallinity, and in turn the electrical properties, of the films depends on the conversion temperature as seen previously for TAC-grown MoS$_2$ films[5]. XPS data for the films converted at different temperatures are shown in the Supporting Information, Figure S5. These suggest a greater contribution from Pt oxides and amorphous Se at the lowest growth temperature used.

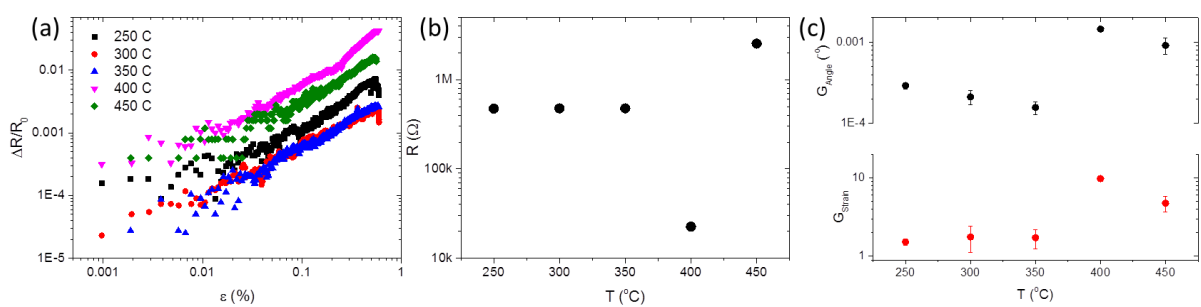

*Figure 3(a) Fractional resistance change as a function of applied strain for 1 nm Pt films on polyimide which were converted to PtSe$_2$ at different temperatures. (b) Measured resistance values for PtSe$_2$ films grown at different temperatures. (c) Extracted gauge factor for PtSe$_2$ films grown from 1 nm Pt at different temperatures.*

The results of cycled testing on a 1 nm Pt film on polyimide, which was converted to PtSe$_2$ at 400 °C, are shown in **Figure 4(a)**. It is clear from this that the electromechanical response is stable over many cycles. Data acquired over a longer timeframe are shown in the Supporting Information, Figure S6. The effect of repeated peeling of the active PtSe$_2$ layer on a PtSe$_2$/polyimide sample (grown from 1 nm Pt) with sticky tape is shown in **Figure 4(b)**. While an increase in the film resistance is observed, this peeling does not have a drastic effect on the electrical properties of the channel suggesting that the PtSe$_2$ adheres very strongly to the polyimide substrate and that the sensors possess good mechanical stability. This strong adhesion to polyimide is noteworthy as structural robustness is paramount if any real-world sensing applications are to be considered. The strong adhesion is also interesting when one considers that previous studies have indicated that PtSe$_2$ can be easily transferred off SiO$_2$/Si substrates[8]. This indicates that the choice of growth substrate has a big influence on the strength of the PtSe$_2$/substrate interaction. The effect of soaking PtSe$_2$/polyimide (grown from 1 nm Pt) in water and acetone for a period of 5 days is shown in **Figure 4(c, d)**. Immersion in water has a minimal effect on the electromechanical properties suggesting that strain gauges based on PtSe$_2$/polyimide could be used in aqueous conditions. Immersion in acetone on the other hand has a severe impact on the electromechanical properties. As polyimide is known to be resistant to acetone the reason for this performance degradation is unclear. Additional detail on these solvent soak tests is presented in the Supporting Information, Figure S7 and S8. Overall, the PtSe$_2$/polyimide strain gauges produced offer good cyclability and stability.

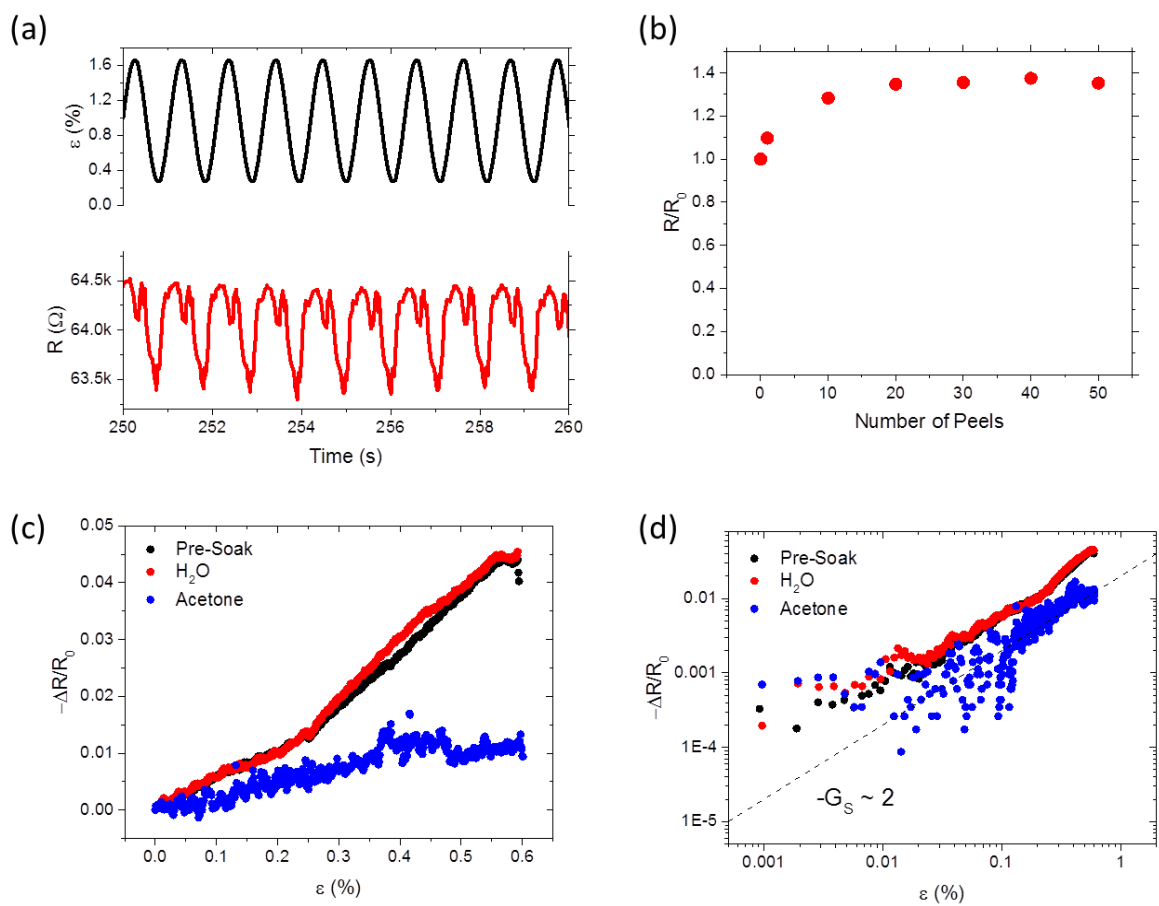

*Figure 4(a) Cycled electromechanical response of PtSe$_2$ on polyimide (from 1 nm Pt). The black line shows the applied strain and the red line shows the change in resistance (b) Fractional resistance change of PtSe$_2$ on polyimide upon repeated peeling. While there is an increase in the resistance the film is still conductive following 50 peels (c, d) Fractional resistance change of reference PtSe$_2$ on polyimide, and films that have been immersed in solvent (water or acetone) for 5 days, under applied strain.*

Preliminary tests were carried out to investigate the response of piezoresistive sensors based on PtSe$_2$ on polyimide to vibrations of different frequencies. This was achieved by attaching a film to a speaker that was then operated at different frequencies. The results of this are shown in Figure 5(a) where a strong signal, with a high signal-to-noise ratio, is seen for vibrations with frequencies of 95 Hz, 190 Hz and 380 Hz. This is further emphasized by the zoomed-in view of the response to vibrations of 380 Hz shown in Figure 5(b). This strong response to high-frequency mechanical vibrations suggests that these sensors could potentially be used to monitor the performance of machinery or other structures with moving/vibrating parts.

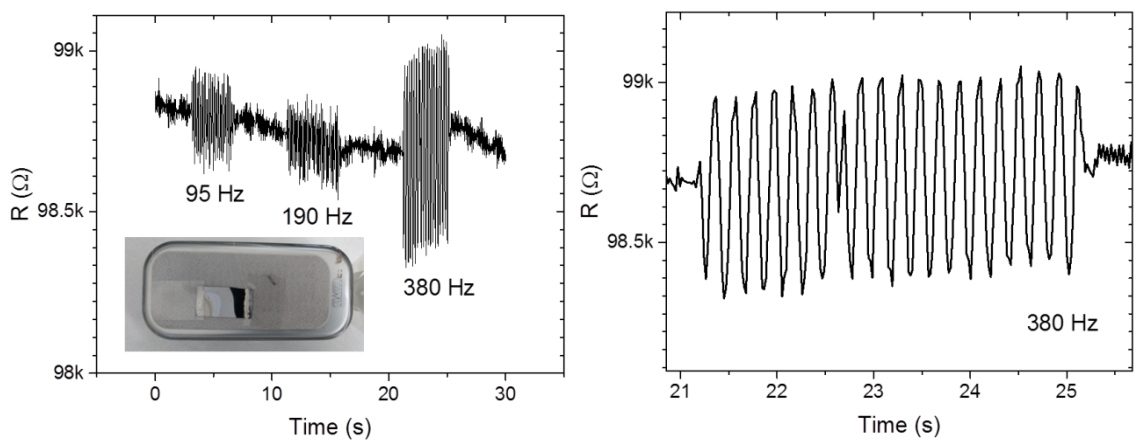

*Figure 5(a) Resistance change of PtSe$_2$/Polyimide attached to a speaker which is operated at different frequencies. Inset: photograph of PtSe$_2$/Polyimide attached to speaker (b) Zoomed-in view of change in resistance of PtSe$_2$/Polyimide when speaker is operated at 380 Hz.*

**Conclusion**

Thin films of PtSe$_2$ have been grown directly on flexible polyimide foil and subsequently assessed for use as strain gauges. These films show superior gauge factors compared to commercial metal-based strain gauges. We have investigated the effect of growth temperature and film thickness on the performance of PtSe$_2$/polyimide strain gauges. The strain gauges were shown to be robust and stable over many cycles. The PtSe$_2$ layer adheres very strongly to the polyimide, with the films remaining conductive after repeated peel tests. The strain gauges show no sign of degradation following immersion in water for prolonged periods. A preliminary investigation showed that the strain gauges are well-suited to high-frequency operation. The combination of sensitivity, stability and robustness means that these strain gauges could potentially be used to monitor repeated vibrations, possibly in an outdoor setting. Lastly, strain gauges are but a very simple example of a technology based on PtSe$_2$ grown on a flexible substrate. We envisage that similar approaches could be developed to fabricate other, more complicated components for flexible electronics.

## 4. Experimental Section

*PtSe$_2$ Growth:* PtSe$_2$ layers were grown by a TAC process as detailed elsewhere previously[7-8]. Pieces of polyimide foil of 125 µm thickness (Upilex 125S, UBE Europe GmbH) were used as substrates. Pt layers of different thickness were deposited on top of the substrates by e-beam evaporation (Temescal FC-2000). These Pt layers were then converted to PtSe$_2$ in a custom-designed, two-zone furnace whereby the Pt/polyimide was heated to a setpoint temperature in zone 1 and Se powder was heated to 250 °C in the upstream zone 2 of the furnace. Forming gas (90% Ar, 10% H$_2$) flow at 150 sccm carried Se vapour from zone 2 to zone 1 of the system. A dwell time of 2 hours was used, previously shown to be sufficient for complete conversion of Pt to PtSe$_2$.

*Characterization:* Atomic force microscopy was performed using a Bruker multimode 8 system. The topographical and conductive mappings were acquired by PeakForce (PF) tapping using the PF-TUNA mode, with standard PF-TUNA probes. The DC bias ranged from 25-100 µV, and the contact current is displayed (the averaged current during the contacted portion of the force curve.)

The XPS data were collected using an Omicron XM1000 MkII X-ray source with monochromatic Al Kα X-rays alongside an Omicron EA125 energy analyzer. An analyzer pass energy of 15 eV was used for the collection of all core-level spectra. An Omicron CN10 electron flood gun was used for charge compensation and the binding energy scale was referenced to the carbon 1s core level at 284.8 eV. Analysis and fitting of the spectral components was performed using CasaXPS software. Spectral components were fitted using a Shirley background subtraction and Gaussian-Lorentzian line shapes. Scanning electron microscopy imaging was done using a Zeiss Ultra Plus SEM operating at 3 kV with an in-lens detector.

*Electromechancial Characterisation*: Electromechanical measurements during flexure tests were performed using a Keithley KE2601 source meter in 2-probe mode, controlled by LabView software, in conjunction with a Zwick Z0.5 ProLine Mechanical Tester (100 N Load Cell). For electromechanical flexure tests, sample specimens were hand-cut with a blade into 20 mm x 17.5mm (L x W) dimensions. Silver wire contacts were painted directly onto the grown $PtSe_2$ layer at each end of the sample and attached to the source meter via crocodile clips.

For the flexure testing, a custom made three-point flexure rig was used. Samples were placed with the $PtSe_2$ side lying face down across two insulting support pins. The mechanical tester driven loading pin was then lowered into its start position ($\varepsilon = 0$) resting against the upward facing substrate side. Samples were then strained at a rate of 10 mm/min until $\varepsilon \sim 0.8\%$.

Bending-beam experiments were conducted by attaching hand-cut samples with super glue (super glue Z70 by HBM) to a steel beam (300 mm long, 30 mm wide and 3 mm thick) at a location 200 mm from the free-standing end of the beam. Masses of 2 kg and 1 kg were attached to the end of the beam resulting in a strain of 0.04 % and 0.02%, respectively. The resistance across the $PtSe_2$ was measured with a Keithley 4200 SCS parameter analyser with 0.1 ms pulses at $\pm 3$ V, $\pm 5$ V and $\pm 7$ V. Measurements were also conducted for tensile (sample at the top of the beam) and compressive (sample at the bottom of the beam) strain.

*Durability Testing*: For all testing, a new $PtSe_2$ layer grown from 1 nm Pt on a 125 μm thick polyimide substrate was used for each test. This sample type was chosen due to its high electromechancial response. For cyclic testing, contacts were painted on the sample. The sample was then placed in the three-point flexure rig and deformed under a sine wave strain output at a frequency of 1 Hz between ε ~ 0.2% and ~ 1.2% for 500 cycles while measuring the electrical response as a function of time. For the tape peel test, clear domestic scotch tape was rubbed on the $PtSe_2$ layer of the sample and peeled off, by hand, 50 times. After each peel, the piece of tape was replaced with a new one to maintain a constant adhesion. The electrical properties of the sample were measured at the $0^{th}$, $1^{st}$, $10^{th}$, $20^{th}$, $30^{th}$, $40^{th}$, and $50^{th}$ peel using a multimeter. For the solvent soak test, three separate samples (one for reference) were soaked in normal tap water and acetone. Every 24 hrs, up to 120 hrs, samples were removed from their respective liquid and oven dried at 50° C for 10 mins. Electrical properties were then measured using a multimeter, after which, the samples were returned to their soak. After the 120 hrs had elapsed, flexure tests were performed on the three samples.

**Supporting Information**

Supporting Information is available online

**Acknowledgements**

N. Mc E. acknowledges support from Science Foundation Ireland (SFI) through 15/SIRG/3329. G. S. D., C. Ó C. and C. P. C. acknowledge the support of SFI under Contract No. 12/RC/2278 and PI15/IA/3131. J. B. Mc M. acknowledges an IRC scholarship under Award 13653. G. S. D and M. C. L. acknowledge the Graphene Flagship under Contract 785219. We thank UBE Europe GmbH for providing polyimide foils.

**ToC**

Thin films of platinum diselenide ($PtSe_2$) are grown directly on flexible polyimide foils. When strain is applied, these films show a strong electrical response. The robustness and stability of the films, combined with the cyclability of the response, suggests that these films could be used in strain gauges and related applications.

**$PtSe_2$, strain gauges, electromechanical sensors, piezoresistivity, flexible electronics, transition metal dichalcogenides**

*Conor S. Boland\*, Cormac Ó Coileáin, Stefan Wagner, John B. McManus, Conor P. Cullen, Max C. Lemme, Georg S. Duesberg and Niall McEvoy\**

**$PtSe_2$ grown directly on polymer foil for use as a robust piezoresistive sensor**

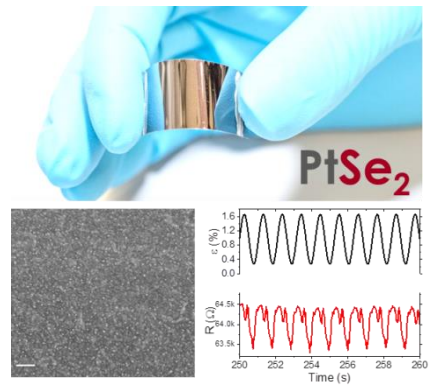

Supporting Information

**PtSe$_2$ grown directly on polymer foil for use as a robust piezoresistive sensor**

*Conor S. Boland\*, Cormac Ó Coileáin, Stefan Wagner, John B. McManus, Conor P. Cullen, Max C. Lemme, Jonathan N. Coleman, Georg S. Duesberg and Niall McEvoy\**

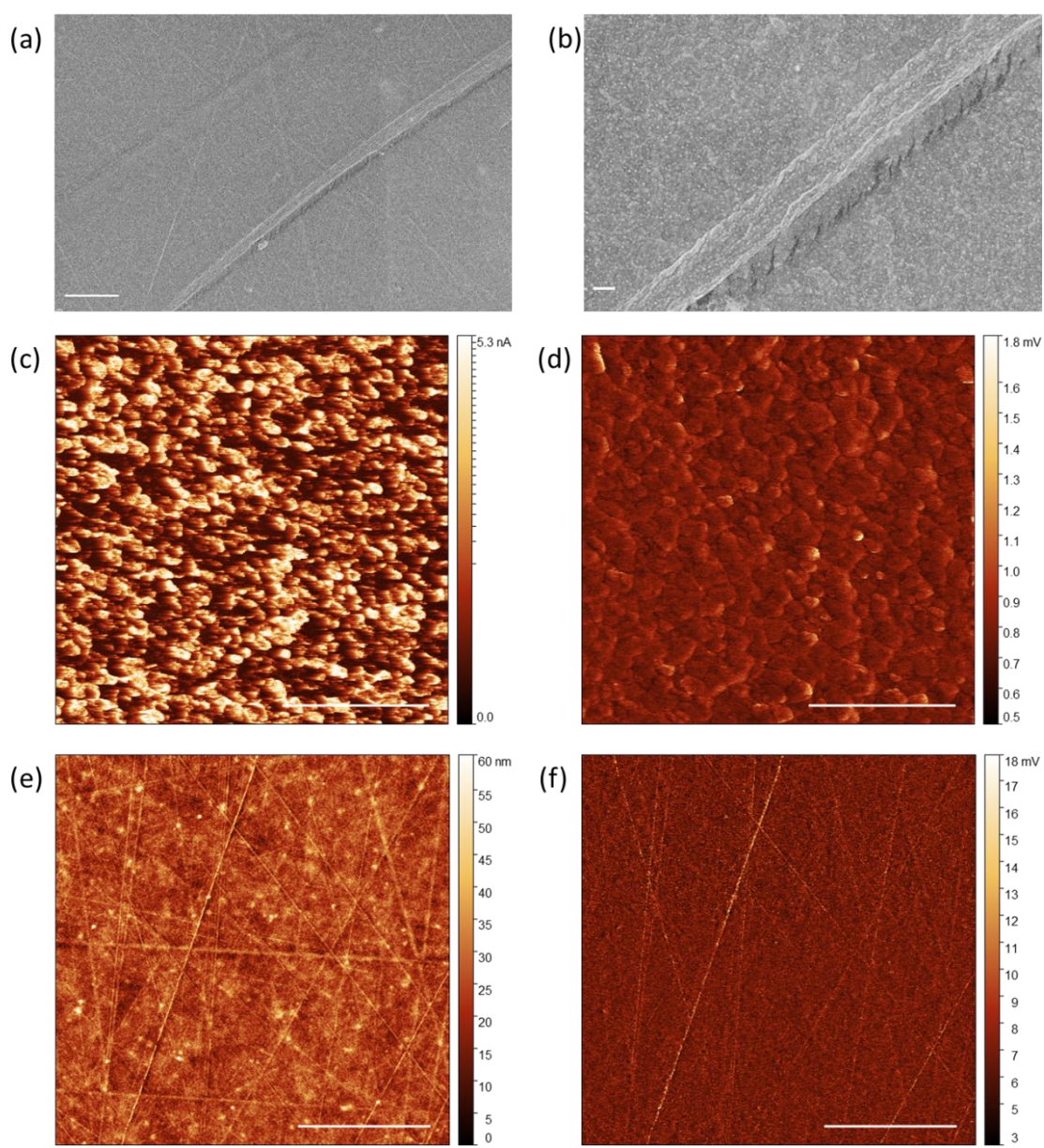

*Figure S1:(a, b) Low-magnification SEM images of PtSe$_2$ from 1 nm Pt on polyimide. Scale bars are 2 μm and 200 nm, respectively. (c, d) Corresponding contact current and deformation AFM images for height image in Figure 1(d) of main text (scale bar = 500 nm). (e, f) Corresponding height and deformation AFM images for contact-current image in Figure 1(e) of main text (scale bar = 10 μm).*

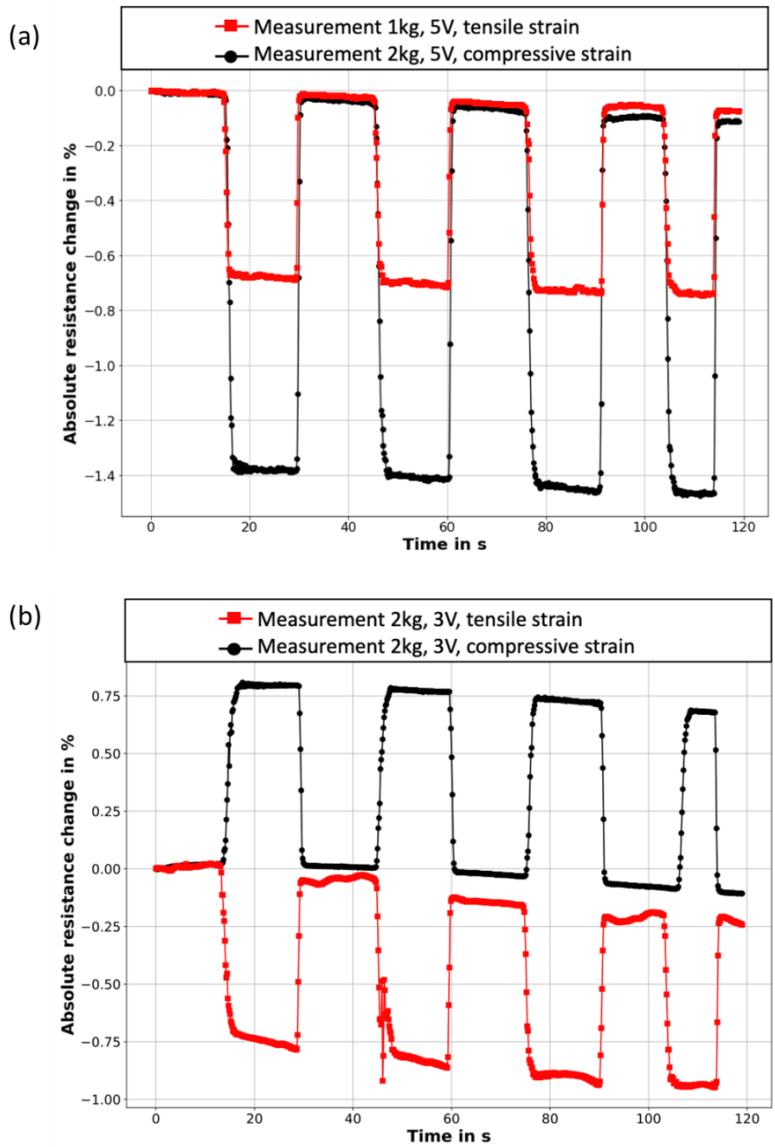

*Figure S2: Cantilever bending-beam measurements. (a) Shown is the resistance change of a PtSe$_2$ on polyimide film grown from 1 nm Pt upon the application of 1 kg and 2 kg masses in a cantilever bending beam setup. This was measured under an applied voltage of 5 V with a cycle duration of 15 s. The extracted gauge factor is ~-30 which is lower than, but of the same order of magnitude as, transferred PtSe$_2$ films previously measured in a similar setup.(b)Measurements of the same film under compressive and tensile strain with a 2 kg mass and an applied voltage of 3 V.*

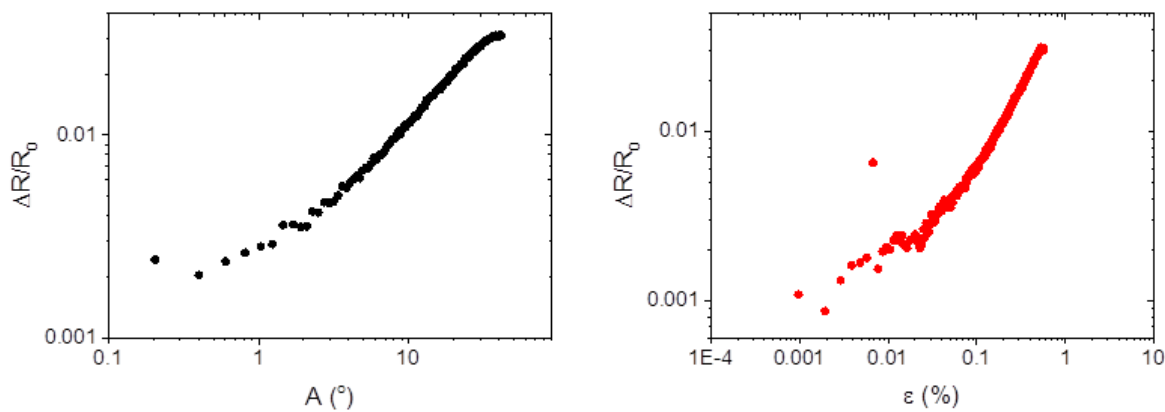

*Figure S3: Electromechanical response of reference 1 nm Pt film on polyimide (not selenised). This shows an increase in resistance under applied strain, leading to a positive gauge factor, as expected for a metal.*

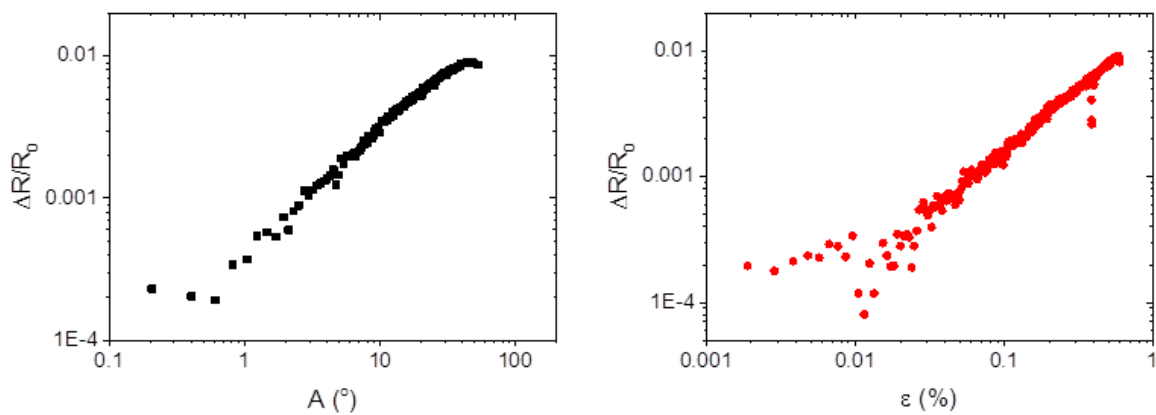

*Figure S4: Electromechanical response of 1 nm Pt film on polyimide which was sulfurised rather than selenised. Under the reaction conditions used the formation of PtS is expected. This shows an increase in resistance under applied strain, giving rise to a positive gauge factor.*

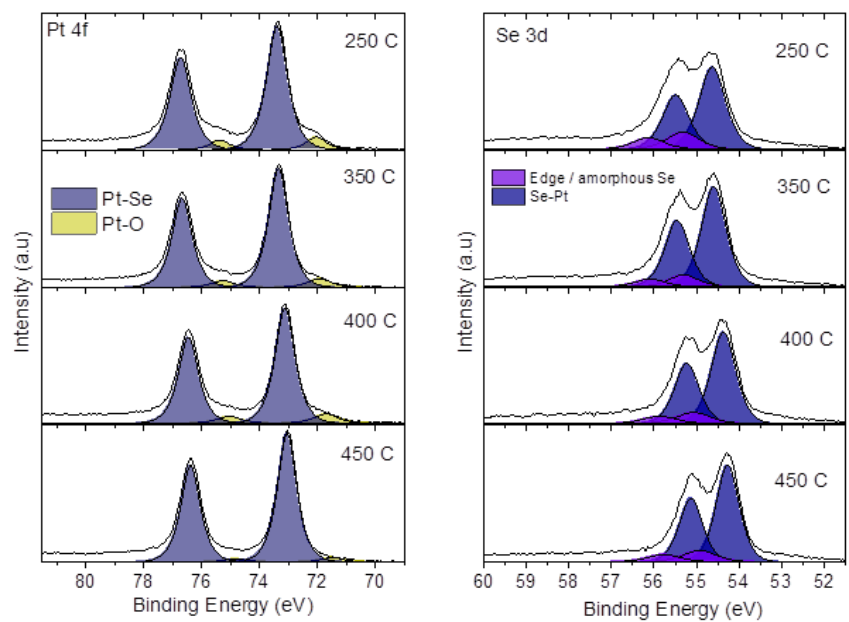

*Figure S5: Pt 4f and Se 3d core level spectra of PtSe$_2$ films grown from 1 nm Pt on polyimide at different temperature.*

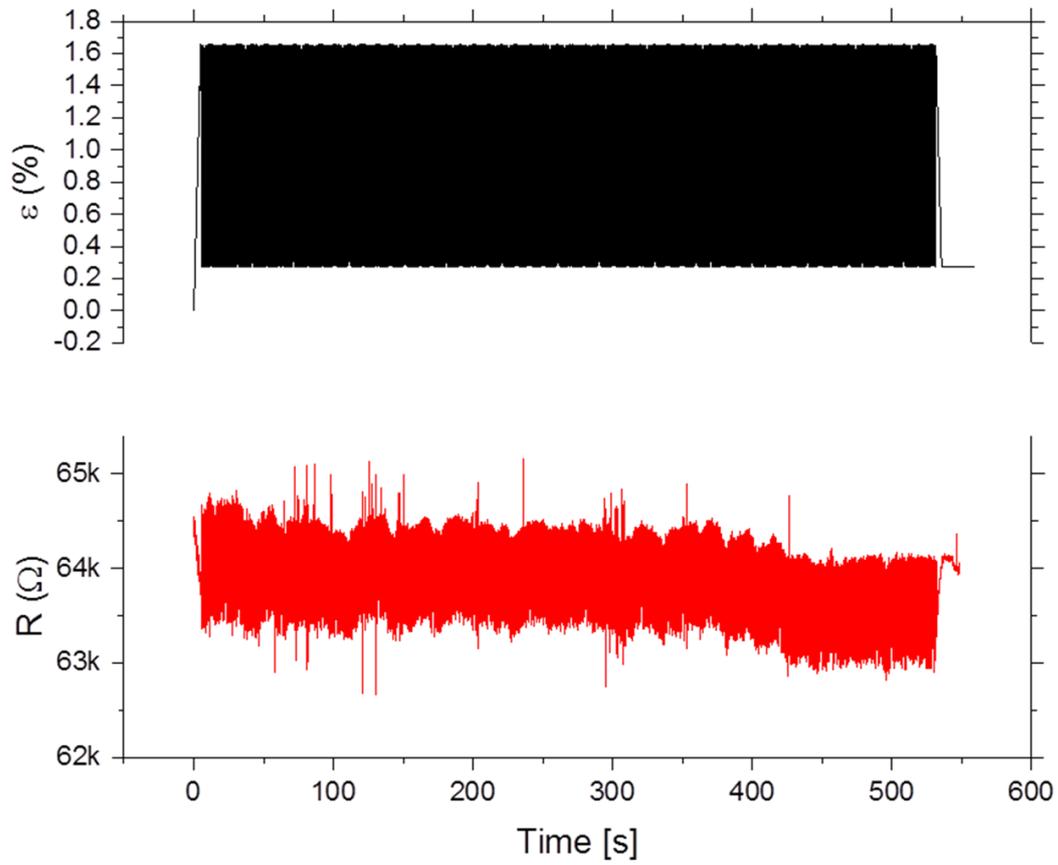

*Figure S6: Repeated cycling of PtSe$_2$ on polyimide (from 1 nm Pt film) shown in Figure 4(a) of main manuscript.*

## Peel Testing

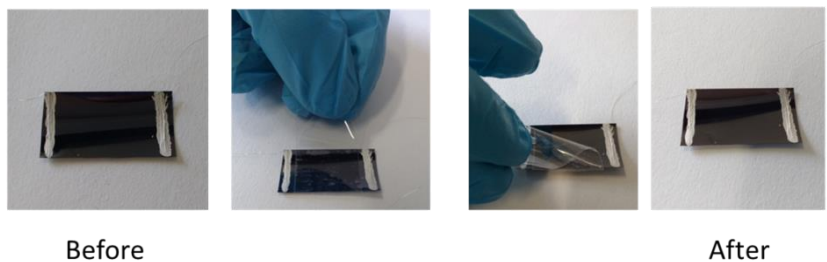

Before                                         After

## Solvent Soak

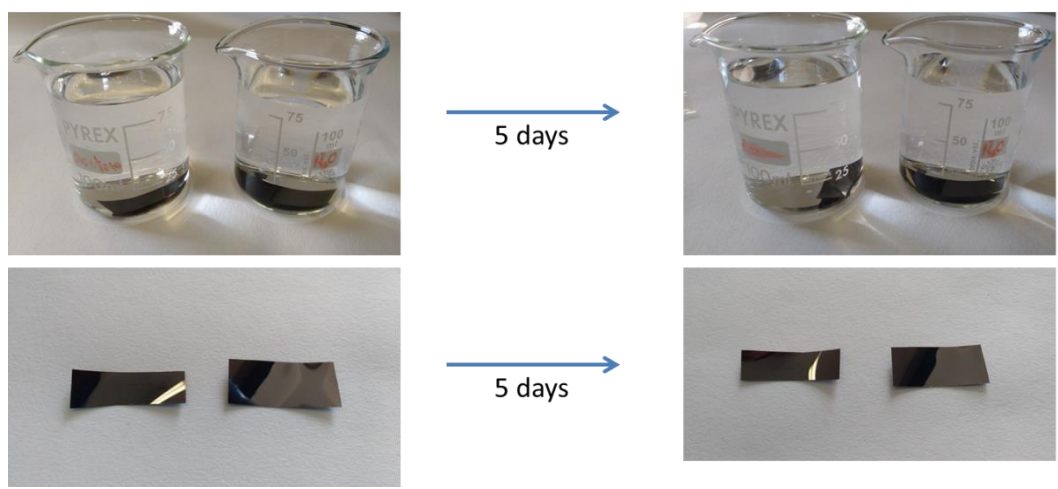

*Figure S7: **Peel testing**: 1 nm Pt thickness film sample has tape applied and peeled off 50 times. Different piece of tape for each peel is used. **Solvent soak**: 1 nm Pt thickness film sample cut into two pieces is soaked in deionised water and acetone for 5 days. Solvent is replenished each day.*

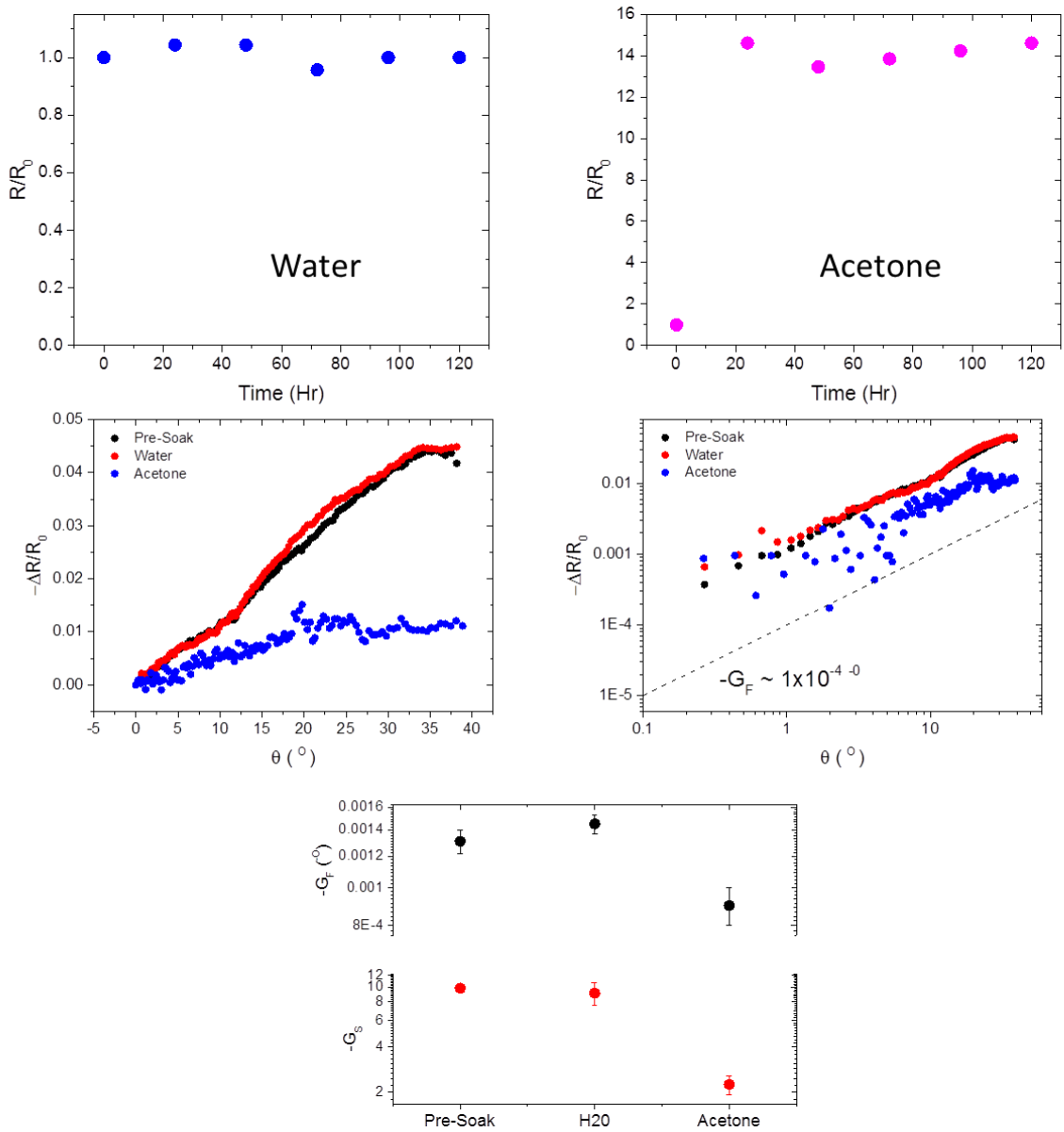

*Figure S8. Additional data for solvent soaked samples. The soaked samples were PtSe$_2$ from 1 nm Pt on polyimide converted at 400 $^o$C (a, b) Change in resistance over time due to immersion in water and acetone. (c, d) Effect of solvent soak on the change in fractional resistance as a function of applied flexure angle. (e) Extracted gauge factors for pristine sample, water-soaked sample and acetone-soaked sample.*